\documentclass[12pt]{article}
\usepackage{epsfig}
\usepackage{psfig}
\advance\textheight by \topskip
\begin{document}
\tolerance=100000
\thispagestyle{empty}
\setcounter{page}{0}
\def\cO#1{{\cal{O}}\left(#1\right)}
\newcommand{\be}{\begin{equation}}
\newcommand{\ee}{\end{equation}}
\newcommand{\br}{\begin{eqnarray}}
\newcommand{\er}{\end{eqnarray}}
\newcommand{\ba}{\begin{array}}
\newcommand{\ea}{\end{array}}
\newcommand{\bi}{\begin{itemize}}
\newcommand{\ei}{\end{itemize}}
\newcommand{\bn}{\begin{enumerate}}
\newcommand{\en}{\end{enumerate}}
\newcommand{\bc}{\begin{center}}
\newcommand{\ec}{\end{center}}
\newcommand{\ul}{\underline}
\newcommand{\ol}{\overline}
\newcommand{\ra}{\rightarrow}
\newcommand{\sm}{${\cal {SM}}$}
\newcommand{\as}{\alpha_s}
\newcommand{\aem}{\alpha_{em}}
\newcommand{\ycut}{y_{\mathrm{cut}}}
\newcommand{\susy}{{{SUSY}}}
\newcommand{\Dir}{\kern -6.4pt\Big{/}}
\newcommand{\Dirin}{\kern -10.4pt\Big{/}\kern 4.4pt}
\newcommand{\DDir}{\kern -10.6pt\Big{/}}
\newcommand{\DGir}{\kern -6.0pt\Big{/}}
  \def\Ecm{\ifmmode{E_{\mathrm{cm}}}\else{$E_{\mathrm{cm}}$}\fi}
\def\gluino{\ifmmode{\mathaccent"7E g}\else{$\mathaccent"7E g$}\fi}
\def\photino{\ifmmode{\mathaccent"7E \gamma}\else{$\mathaccent"7E \gamma$}\fi}
\def\mgluino{\ifmmode{m_{\mathaccent"7E g}}
             \else{$m_{\mathaccent"7E g}$}\fi}
\def\taugluino{\ifmmode{\tau_{\mathaccent"7E g}}
             \else{$\tau_{\mathaccent"7E g}$}\fi}
\def\mphotino{\ifmmode{m_{\mathaccent"7E \gamma}}
             \else{$m_{\mathaccent"7E \gamma}$}\fi}
\def\ML{\ifmmode{{\mathaccent"7E M}_L}
             \else{${\mathaccent"7E M}_L$}\fi}
\def\MR{\ifmmode{{\mathaccent"7E M}_R}
             \else{${\mathaccent"7E M}_R$}\fi}
\def\lsim{\buildrel{\scriptscriptstyle <}\over{\scriptscriptstyle\sim}}
\def\gsim{\buildrel{\scriptscriptstyle >}\over{\scriptscriptstyle\sim}}
\def\MCH {$\tilde\chi_1^+$}
\def \CH{{\tilde\chi}^{\pm}}
\def \LSP{\tilde\chi_1^0}
\def \SNU{\tilde{\nu}}
\def \BARSNU{\tilde{\bar{\nu}}}
\def \MLSP{m_{{\tilde\chi_1}^0}}
\def \MCH{m_{{\tilde\chi}^{\pm}}}
\def \MCHMIN {\MCH^{min}}
\def \ET{\not\!\!{E_T}}
\def \LL{\tilde{l}_L}
\def \LR{\tilde{l}_R}
\def \MLL{m_{\tilde{l}_L}}
\def \MLR{m_{\tilde{l}_R}}
\def \MSNU{m_{\tilde{\nu}}}
\def \PI{{\pi^{\pm}}}
\def \DM{{\Delta{m}}}
\newcommand{\bQ}{\overline{Q}}
\newcommand{\ad}{\dot{\alpha }}
\newcommand{\bd}{\dot{\beta }}
\newcommand{\dd}{\dot{\delta }}
\def \CH{{\tilde\chi}^{\pm}}
\def \MCH{m_{{\tilde\chi}_1^{\pm}}}
\def \LSP{\tilde\chi_1^0}
\def \MUL{m_{\tilde{u}_L}}
\def \MUR{m_{\tilde{u}_R}}
\def \MDL{m_{\tilde{d}_L}}
\def \MDR{m_{\tilde{d}_R}}
\def \MSNU{m_{\tilde{\nu}}}
\def \MLL{m_{\tilde{l}_L}}
\def \MLR{m_{\tilde{l}_R}}
\def \mhf{m_{1/2}}
\def \MST{m_{\tilde t_1}}
\def \lum{{\cal L}}
\def \RPVC{\lambda'}
\def\tth{\tilde{t}\tilde{t}h}
\def\qqh{\tilde{q}_i \tilde{q}_i h}
\def\t1{\tilde t_1}
\def \pt{p{\!\!\!/}_T} 
\def \etm{E{\!\!\!/}_T} 
%#############################
%#$\tan\beta \approx 3$ 
\def\lapp{\mathrel{\rlap{\raise.5ex\hbox{$<$}}
                    {\lower.5ex\hbox{$\sim$}}}}
\def\gapp{\mathrel{\rlap{\raise.5ex\hbox{$>$}}
                    {\lower.5ex\hbox{$\sim$}}}}
\newcommand{\decay}[2]{
\begin{picture}(25,20)(-3,3)
\put(0,-20){\line(0,1){15}}
\put(0,-20){\vector(1,0){15}}
\put(0,0){\makebox(0,0)[lb]{\ensuremath{#1}}}
\put(25,-20){\makebox(0,0)[lc]{\ensuremath{#2}}}
\end{picture}}
%#############################
\vspace*{\fill}
\vspace{-1.2in}
\begin{center}
%{\Large DRAFT v 1.7}\\
\end{center}
\begin{flushright}
{DESY 03-042}\\
{SHEP-03-06}\\
%{\today}\\
{TIFR/EHEP/03-03}\\
{TIFR/TH/03-08}\\
{hep-ph/0304137}\\
\end{flushright}
\begin{center}
{\Large \bf
Search for `invisible' Higgs signals at LHC \\[0.25cm]
via Associated Production with Gauge Bosons
}\\[0.3cm]
\end{center}
\begin{center}
{\large R.M. Godbole${^{a,}}$\footnote{Permanent Address: 
Centre for Theoretical Studies, IISc, Bangalore, 560 012, India}, 
M. Guchait${^b}$, K. Mazumdar${^{c,}}$\footnote{Permanent Address:
Department of High Energy Physics, TIFR, Mumbai, 400 005, India},\\[0.20cm]
S. Moretti$^d$ and D.P. Roy$^{e,f}$ }\\[0.25 cm]
{\it $^a$ Theory Division, DESY, Notkestrasse 85, D-22603 Hamburg, 
Germany\\[0.20cm]
$^b$Department of High Energy Physics, Tata Institute of 
Fundamental Research,\\
Homi Bhabha Road, 400 005 Mumbai, India\\[0.20cm]
$^c$ CERN, EP Division, Geneva, CH-1211, Switzerland\\[0.20cm]
$^d$ Department of Physics \& Astronomy, Southampton University, \\
Highfield, Southampton SO16 7GH, UK\\[0.20cm]
$^e$ Department of Theoretical Physics, Tata Institute of Fundamental
Research,\\
Homi Bhabha Road, 400 005   Mumbai, India\\[0.20cm]
$^f$ Universitat Dortmund, Institut fur Physik,\\
D-44221 Dortmund, Germany}
\end{center}

\vspace{.4cm}

\begin{abstract}
{\noindent\normalsize 
A light Higgs boson 
with substantial branching ratio into invisible channels
can occur in a variety of models with: light neutralinos, 
spontaneously broken lepton number,
radiatively generated neutrino masses, additional singlet scalar(s) 
and/or right 
handed neutrinos in the extra dimensions of TeV scale gravity. We study the 
observability of the $W H$ 
and $ZH$ modes at LHC with $H$ decaying invisibly,
by carrying out a detailed simulation with two
event generators ({\tt HERWIG} and {\tt PYTHIA}) and realistic  
detector simulations (GETJET and CMSJET). We find that the signal 
with `single lepton plus missing $E_T$'
resulting from $W H$ production suffers from a very large background 
due to the (off-shell) $W^{*}$ production 
via the Drell-Yan process. In contrast, the $ZH$ mode provides
a clean signal in the `dilepton plus missing $E_T$' channel.
By exploiting this second signature, we show that
invisible branching ratios
of Higgs bosons, ${\mathrm{{BR}_{~inv}}}$, 
larger than $\sim 0.42(0.70)$ can be probed at 5$\sigma$ level for 
$M_H = 120$(160) GeV respectively, assuming an accumulated luminosity of
${\cal L}=100$ fb$^{-1}$.}

\end{abstract}
%PACS no: 12.60 Jv,13.38Be
\vskip1.0cm
\noindent
\vspace*{\fill}
\newpage
\section*{Introduction}
%\label{sec_intro}

The search for Higgs bosons and the study of their properties is one of 
the main goals of physics studies at Tevatron upgrade (Run 2) and the
upcoming Large Hadron Collider (LHC). 
The precision measurements with Electro-Weak (EW) data indicate the 
existence of a light Higgs boson ($M_H <204$~GeV at 95\% C.L.) 
whereas direct searches rule out the case
$M_H < 114.4$ GeV \cite{hagiwara}, also giving a hint of a possible signal at 
the very upper end of the experimentally excluded interval \cite{lep}.
Naturalness arguments along with the indication of a light Higgs 
state suggest that Supersymmetry (SUSY) is a likely candidate for new physics 
Beyond the Standard Model (BSM). In most SUSY scenarios, 
the Lightest SUSY Particle (LSP) is the neutral, weakly interacting and stable
neutralino, denoted as
$\LSP$. The current combined limits on the neutralino and Higgs 
boson masses in a general SUSY model \cite{susylim} are such that, for 
non-universal gaugino masses
at high scale, it is still kinematically possible for a
relatively light Higgs state to decay
into two LSPs with a large Branching Ratio (BR), as high
as 0.70, without being in conflict with the relic density and the 
$(g-2)_{\mu}$ constraints \cite{godbole}. In such a case, the Higgs boson 
becomes {\it invisible}. 
Other models of invisible Higgs decay are 
connected to possible scenarios for
neutrino ($\nu$) mass generation.
One of the mechanisms for the latter arising in theories with extra 
dimensions and TeV scale gravity \cite{arkani}, for example, can 
cause the $H$ to have several invisible decay modes. Here, 
$H$ states can decay into 
$\nu_L \bar\nu_R^j$ where $\bar \nu^j_R$ denotes the $j$th Kaluza-Klein 
(KK) excitation of the light neutrino which is a singlet. The tall tower
of KK resonances can cause the width $\sum_j 
\Gamma( H \ra \nu_L \bar\nu_R^j)$ to be sizable. Besides, in models
where a Majorana mass of $\nu$'s results from a spontaneously broken global 
symmetry \cite{joshipura}, $H$ states can have appreciable branching 
fractions into  two Nambu-Goldstone bosons.
This type of decay mode may also arise in some models with 
extended higgs sector with an additional higgs singlet in the framework of 
Standard Model(SM)~\cite{arc}.
Finally, if the neutrino mass is generated radiatively by some mechanism
 below the TeV scale,
again a Higgs boson may decay invisibly into a $\nu_{\rm{light}} 
\nu_{{\rm{heavy}}}$ pair \cite{pilaftsis}. Similarly, Higgs boson may
also decay invisibly 
into a pair of neutrinos in the framework of models with 4th generation 
lepton~\cite{4thnut}.

Needless to say, the only
possible mode in which a Higgs boson can decay invisibly in the
SM is via $H\to ZZ^\ast \to 4\nu$ 
which has a BR of about 1\% at $ M_H>$ 180 GeV
and even lower at lower values of $M_H$. Thus it can not disturb
the visible Higgs decay modes appreciably for any value of $M_H$.
On the other hand in the above mentioned BSM scenarios the 
invisible decay mode can represent a large
part of the decay BR for an Intermediate Mass Higgs (IMH) boson,
with 114~GeV$ \lsim M_H \lsim $160~GeV.
In this mass range, in fact, detection
of a Higgs signal relies mainly on the $b\bar b$, $\gamma\gamma$,
$W W\to 2\ell ~2\nu_\ell$ and $ZZ^\ast \to 4\ell$ final states,  
but only as long as the corresponding rates are not very 
different from the SM values. A reduction of the latter, due to the presence
of sizable invisible decays of an IMH boson, could prevent its detection
at Tevatron and LHC. 

In short, in a large number of BSM physics scenarios, all addressing the
fundamental issues of $\nu$ mass generation and/or radiative stability of 
the EW symmetry 
breaking scale, there exists the possibility of an IMH boson decaying  into 
invisible channels, thus hindering the chances of its discovery via
the customary decay modes studied so far that we have listed above. Hence, it is
necessary to develop new search strategies for these otherwise lost
signal events, in order to confirm 
the ability of present and future colliders to resolve the structure of 
 the Higgs sector. 

%---------------------------- New Section -----------------------------% 
\section*{Leptonic Signatures of Invisible Higgs Decays}

Since Tevatron is in operation at present and LHC is the 
next world machine, it is natural to review the current
status of invisible Higgs decays starting from the case of hadronic 
colliders\footnote{In 
the more distant future, leptonic colliders may be of some help
in extracting signatures  of invisibly decaying Higgs bosons. For
example, at a TeV scale $e^+e^-$ linear collider (LC), invisible Higgs
decays should be readily accessible via 
 $e^+e^- \rightarrow Z H \ra (Z \ra b \bar b) 
( H \ra $~invisible)
\cite{campos}.}.
A study \cite{martin} of the possibilities at Tevatron Run 2 has
shown that, even with an integrated luminosity of 30 fb$^{-1}$, evidence of
an invisibly decaying Higgs boson is possible only at 3$\sigma$ level and 
no further than
$M_H=135$ GeV. A 5$\sigma$ discovery for an $M_H$ value beyond the LEP limit
will require a luminosity as high as 50--70 fb$^{-1}$, which is unattainable
at Tevatron. At LHC there is more scope because of the 
higher luminosity  as well as the much larger Higgs boson production cross 
sections.
Here, the dominant (at least in a SM scenario \cite{review}) Higgs production
process is gluon-gluon fusion via a top-quark loop  ($gg\to H$). 
If the Higgs boson decays invisibly, the only way the
$gg$ channel can be used is by looking at the production of
$H$ in association with a jet, i.e.,
 $gg \ra H + {\mathrm {jet}}$.  The signature 
here would be events with modest amount of missing (transverse) energy and 
a large $E_T$ 
jet, the `monojet' events, in presence of underlying hadronic activity.  This 
signal is however overwhelmed by the pure QCD background, so that the 
potential of this channel for Higgs detection 
is very limited already at the parton level \cite{choudhuri}. 

It is then natural to turn to the subleading Higgs production channels. 
These are in turn dominated by vector-vector fusion ($qq\to qq V^*V^*\to qq H$), 
followed by
Higgs-strahlung ($q\bar q^{(')}\to VH$)\footnote{Here, the labels 
$\bar q^{(')}$ refer
to (anti)quarks of any possible flavour
whereas $V=W,Z$.} and associate production with
top-quarks ($gg\to t\bar tH$). The vector-boson fusion process suffers from
the same drawbacks of the leading channel, as the final signature
would again be purely hadronic. Nonetheless, recent studies~
\cite{eboli} have shown that simultaneous forward and backward jet tagging along with 
rejection of central jets
in the $VV$ fusion mode might provide possibilities of 
detecting an invisible Higgs, for a
${\mathrm{{BR}_{~inv}}}$ value as low as 5\%, with 100 fb$^{-1}$ luminosity. 
In contrast, the last two production modes
naturally offer the possibility of high $E_T$ electron/muon tagging,
by exploiting the leptonic decays of the vector bosons and top-quarks
respectively,
thus providing an effective handle against the pure QCD backgrounds. Previous 
studies of both the Higgs-strahlung production modes, $ZH$ and $WH$, 
followed by the leptonic decays of $W/Z$ \cite{choudhuri,reid}, showed 
that this channel can be efficient for ${\mathrm{{BR}_{~inv}}} \gsim 25$\% at 
100 fb$^{-1}$, whereas the $t \bar t H$ mode would require 
${\mathrm{{BR}_{~inv}}} \gsim 60$\% at the same luminosity \cite{gunion}.
Notice that all such studies are however in need of more rigorous 
analyses, as all of them  have been carried out only at the parton level. 
Full simulations, in presence of parton shower, hadronisation and
detector effects, with varying degrees of sophistication, are currently 
in progress for $VV$ fusion \cite{vvsim}, $VH$ \cite{vhsim}
and $t \bar t H$ production \cite{tthsim}. For example, the ATLAS studies 
of the $VV$ fusion case find a 
sensitivity to $\mathrm{{BR}_{~inv}} \gsim 25$\%, for
$M_H = 120$ GeV with an integrated luminosity of $30$ fb$^{-1}$. CMS expects to probe upto $\mathrm{{BR}_{~inv}} \sim 12.5$\% in the same mass region with $10$ fb$^{-1}$. Note both studies have assumed central jet veto survival probabilities obtained from parton level simulation at the level
of NLO.
However these studies have not taken into account the potentially serious
background from diffractive scattering.

The question naturally arises now
whether one can continue to use the SM values
for the $HVV$ couplings controlling the $W H$ and 
$ZH$ production cross sections of our 
interest, while probing BSM scenarios with a large  invisible decay rate of 
an IMH boson. In the BSM scenarios associated with neutrino mass generation, 
the SM values of the $HVV$ couplings are generally compatible with large 
Higgs BRs 
into invisible decay channels~\cite{arkani,joshipura,pilaftsis}. 
In contrast,
in the Minimal Supersymmetric Standard Model (MSSM), with or without 
high scale universality, the SM values of the $HVV$ couplings are suppressed 
by the factor 
$\sin(\alpha-\beta)$, where $\alpha$ and $\beta$ are the mixing angles 
between the two doublets in the neutral and charged Higgs sectors.
It is well known that $\sin(\alpha-\beta) \simeq$1 if the pseudoscalar
Higgs mass is in the range 
$M_A \gsim$ 120 GeV~\cite{Abound}. The LEP lower limit on this mass
is $M_A > 90$ GeV if $\tan\beta \gg$1 and much larger when $\tan\beta \gsim$1.
Thus, except for a tiny slice of the allowed $(\tan\beta,M_A)$ 
parameter plane, i.e., $M_A=90$--115 GeV, the SM values for the 
$HVV$ couplings can be used for the MSSM as well. 

We may add a few general comments here regarding the Higgs production
cross-sections in the MSSM.  While the 
strength of the $VVH$ couplings may be taken as the SM value for 
most of the MSSM parameter space, the presence of relatively light 
squarks and gluinos (i.e., below 1 TeV) can affect the $gg$ fusion
channel and also induce
more production modes for Higgs bosons than those considered
so far. For example, while light squarks may induce cancellation
effects against the quark loops in gluon-gluon fusion \cite{djouadi},
an abundant generation of Higgs states would result
from squark and gluino decays \cite{guchait} and/or associated 
production with squark pairs \cite{sridhar}--\cite{sqsqH}.
It has also been pointed out that $H \tilde\chi_i^0~(i >1$) 
production is sizable and it is maximal \cite{belanger}
where invisible Higgs decays are large. Moreover, for $\tan\beta 
\gsim7$, Higgs production in association with
bottom-quark pairs becomes dominant over the gluon-gluon fusion mode 
and the top-loop contribution to the latter is overwhelmed by
the bottom one.

For the Higgs-strahlung process of our interest
it will then be adequate to simply use the SM production
rates for $q\bar q\to VH$, allow for the Higgs scalar to go undetected,
whatever its final decay products, and sample the detectable 
${\rm BR_{\rm{~inv}}}$ 
values that would allow for the signal extraction above purely SM
backgrounds. 
The model independent approach chosen here is 
sufficiently simple to cover most of the MSSM as well as all the other 
BSM scenarios 
that we have described above. 
In short we will only consider the associated production processes $q \bar q
\ra W H$, followed by the leptonic decay $W \ra \ell \nu$ 
(hereafter, $\ell=e,\mu$),
giving rise to a `single lepton + $E{\!\!\!/}_T$' signature, 
plus $q \bar q \ra Z H$, followed by the leptonic decay
$Z\to \ell\bar\ell$, giving rise to  a `dilepton + $E{\!\!\!/}_T$' signal. 
We shall restrict our analysis  to the case of LHC.

%----------------------New section --------------------------------%
\section*{Kinematic Properties of Signal and Background}
We begin by considering the signal for $WH$  production  coming from 
the process
\be
q \bar q' \rightarrow W^{{ *}} \ra \decay{W~~~}{\ell \nu_{\ell}} ~~~ + ~~~
\decay{H~~~~}{\rm {invisible}}\\
\vspace{2cm}
\label{whsignal}
\ee
This will result in events with a single high  $E_T$ lepton and 
$\pt$\footnote{A subleading contribution to the signal will also 
come from $ZH$ production, with $Z\to \ell\bar\ell$, when one of 
the leptons is lost beyond the lepton detection region 
(typically, $|\eta|\le 2.5$).}.
Since at the parton level $E_T^\ell = \pt$, demanding a large $E_T$ lepton 
automatically ensures a large $\pt$ value. This also means that
one has essentially only one four-momentum at disposal for kinematic
cuts.  Leading backgrounds to the signal are the following with 
$\ell=e,\mu$.
\begin{enumerate}
\item[a)] Charged Drell-Yan (DY)
 production via $q \bar q' \ra W^{{(*)}} \ra \ell \nu_\ell$. 
\item[b)] The irreducible background $q \bar q' \ra W Z \ra 
(\ell \nu_\ell) 
(\nu_{\ell'} \bar \nu_{\ell'})$, with $\ell'=e,\mu, \tau$.
\item[c)]$q \bar q \ra W W \ra (\ell \nu_\ell) (\ell' \nu_{\ell'})$,  
where one of the leptons lies outside the fiducial volume.
\item[d)]$q \bar q' \ra W(\to\ell\nu_\ell) 
+ {\mathrm{jet}}$, when the jet is not detectable due its low $E_T$ or passes through detector cracks and then the lost jet would add onto the missing $E_T$ of the decay $\nu$  from the $W$.
\item[e)]$q\bar q\ra Z 
+ {\mathrm {jets}}$ production with $Z \ra \nu \bar \nu$ can 
also give a background if one jet is misidentified as lepton.
\item[f)]  $q \bar q, gg \ra t \bar t \ra W W b \bar b 
\ra \ell \nu_\ell 
\ell' \nu_{\ell'} b \bar b$($\ell \nu_\ell q \bar q' b \bar b$), 
which may mimic the signal if the $b$-jets are lost along with 
one of the decay leptons(the $W$ decay jets). 
\end{enumerate}

The level of jet activity in background f), coming from hadronic decays of 
$W$ bosons and/or high $E_T$ $b$-quarks, helps to 
distinguish this process from the signal in eq. (\ref{whsignal}). In fact,
the hadronic activity in the signal as well as the purely leptonic background processes mentioned above comes entirely from initial state radiation which is mostly in the forward-backward region of the detector.  Hence, a veto on central jet activity
can help handle $t \bar t$ production effectively. With an expected 
rejection factor of $10^{-5}$ against jet misidentification, the background 
in e) will not be a serious one in the end. 

A useful kinematic variable is the transverse mass, defined as:
\begin{equation}
M_{T}^{n\ell} 
%= \sqrt{ (|\pt| + |E_T^{n\ell}|)^2 - |\pt +E_T^{n\ell}|^2}
 =  \sqrt{2 E_T^{n\ell} \pt (1-\cos\phi(E_T^{n\ell},\pt))},
\label{mtrans}
\end{equation}
where $n=1$ for the single lepton channel and $n=2$ for the dilepton one, 
respectively. In the latter case, $E_T^{2\ell}$ refers to the transverse
component of the three-momentum of the dilepton system.
 
Demanding that $M_{T}^{1\ell} > 100 $ GeV can remove the background 
coming from 
a (real) $W$ in a) without any effect on the signal and the irreducible 
background in b). Unfortunately, it can not suppress the contribution 
coming from a
(virtual) $W^{*}$ in a), which was not  considered in 
\cite{choudhuri}. In fact, while
the sizes of the expected cross sections for $W H$ and $W Z$ 
production are similar for the $M_H$ values under consideration, 
the $W^{*}$
contribution is much larger in comparison. Besides, both the signal
and the DY background are generated via the same 
$s$--channel annihilation, preventing one from exploiting angular
distributions of the visible lepton, in order to enhance the
signal-to-background ratio. One noticeable difference would
be a somewhat broader $E_T^\ell$ (or equivalently the $\pt$ at the parton 
level) distribution for the $W^{{*}}$ background than for the signal 
or the $W Z$ background. So, 
one could imagine choosing a window in the $E_T^\ell$
(or $M_{T}^{1\ell}$) spectrum to handle the off-shell contribution in a). 
As we will see later, this is of too little help to suppress 
$W^{{*}}\to\ell\nu_\ell$ events from a), so that the single lepton channel
will in the end  prove to be unusable. Other possible backgrounds that we
consider are those coming from $ZZ$ and single top production. These
however have a very  small  event rate to start with and do not need any
special kinematic treatment.

The signal for $ZH$ production comes from the process:
\be
q \bar q \rightarrow Z^{{*}} \ra \decay{Z}{\ell\bar\ell} \hspace{.4cm} 
+ \hspace{0.7cm} \decay{H}{\rm {invisible}}
\label{zhsignal}
\vspace{1.5cm}
\ee
This gives rise to a dilepton + $\pt$ signature in the final state. 
The $ZH$ production rate, though a factor of $\sim 5$--$6$ smaller than 
that of $W H$, is more suitable for our search.  
The main backgrounds here are the following.
\begin{enumerate}
\item[x)] DY production of  $Z^{{*}} \rightarrow \ell\bar\ell$ in presence of
jets when the latter get lost.
\item[y)] Irreducible $q \bar q \ra Z Z$ production followed by an
invisible decay of one $Z$ (i.e., $Z \rightarrow \nu_{\ell'} \bar \nu_{\ell'}$), 
with the other $Z$ decaying into a $\ell \bar \ell$ pair. 
\item[z)]  $q \bar q' \ra W Z$ followed by the  leptonic decays of both the  
$W$ and $Z$, giving rise to $ \ell \nu_{\ell} \ell'\bar \ell'$, 
where one lepton is lost. 
\item[w)] $q \bar q' \ra W W $ production followed by  leptonic decays 
of both the $W$'s.
\item[v)]  $q \bar q, gg \ra t \bar t \ra W W b \bar b \ra \ell \nu_\ell 
\ell' \nu_{\ell'} 
b \bar b$, which can cause a background if  
the $b$-jets escape detection.
\end{enumerate}
For the dilepton signal one can demand a large $E_T$ lepton {\it and} 
a high threshold for $\pt$. The latter will largely remove both the on- and 
off-shell components of the background in x) and to a smaller extent the 
background in z). The additional  requirement that the $\ell\bar\ell$ 
mass reconstructs  to $M_Z$ will strongly reduce backgrounds w) and v).
As usual, a veto on the accompanying hadronic activity
in the event further helps to remove the $t \bar t$ background. The only 
limiting factor will be seen to be the irreducible 
background coming from  $ZZ$ production.
The $\pt$ and $E_T^\ell$ 
distributions are softer for the irreducible background than for the signal
as the $ZH$ production is an $s$-channel 
process and the $ZZ$ production occurs via light quark exchange in the 
$t,u$-channel. 
Bearing in mind a possible misidentification of a jet as a lepton, 
we also consider the contributions from
$W H$, $W$ production via DY and single top 
production. These will be however seen to be negligible.
We also checked the background where a final state radiation off a Z
boson of a neutrino--anti-neutrino pair produced via DY process.
The production cross section corresponding to this process, 
$pp \ra \nu \bar\nu Z$ turns out to 
be only 27~fb at LHC, going down to 0.26~fb after the cuts discussed
below. Therefore, it does not appear to be a serious background 
to our signal.   

%-------------------- New Section ------------------------------%
\section*{Simulation and Results}

In our simulation, we have used two different Monte Carlo (MC) event 
generators, {\tt HERWIG} \cite{herwig} and {\tt PYTHIA} \cite{pythia}, 
for comparison. In the two cases, the default settings
of v6.4 and 6.2 (respectively) were adopted.   
While using {\tt HERWIG}, we have adopted {\tt GETJET} \cite{GETJET} 
for calorimeter emulation and jet reconstruction, whereas 
in conjunction with {\tt PYTHIA} we have used {\tt CMSJET} \cite{cmsjet} to 
simulate the detector response specific to the CMS experiment. All the 
possible 
decay modes for the particles generated in the hard scattering process
have been considered and finite width effects have been included for all
the unstable particles with the exception of the top (anti)quark. 
This procedure 
thus includes $\tau$-decays for $W$'s and $Z$'s. However, as is 
clear from the previous sections,
we only consider as signals those involving an $e$ and/or $\mu$ trigger. 
Thus, hadronic $\tau$-decays are typically discarded while leptonic
ones do enter our samples of single and double lepton events, with little
effect in both cases, though. We have used BR($W \to\ell \nu_\ell$)
= 22\% and BR($Z\to\ell\bar\ell$)=6.6\%. In total, we have generated
 $10^6$ MC events for 
each channel in Tabs. 1--2 and processed them through our selection cuts
listed below\footnote{We have always found consistent results between
{\tt HERWIG} and {\tt PYTHIA}, with the only exception of the $W^+W^-$
process, where differences as large as 50\% emerged in the
case of the single lepton analysis but not for the dilepton case. 
We have not been
able to fully understand the discrepancy. However, 
as this background is subleading and the $\ell +\pt$ signal will
be shown to be unviable in any case, we have not pursued this matter further.
In the remainder of the paper, we will only present results from
{\tt HERWIG}.}.

For the  $\ell + \pt$ channel we have enforced the following 
`preliminary' acceptance constraints.
\begin{enumerate}
\item[1a.] Select only one lepton with
$E_T^\ell > 10~{\rm {GeV}} $ and $ |\eta^\ell| < 3.$

\item[2a.] Impose a hadronic veto, by rejecting any events containing jets with
$E_T^j > 30~{\rm {GeV}} $ and $ |\eta^j| < 4.$

\item[3a.] Enforce a missing transverse momentum threshold:
$p{\!\!\!/}_T> 30~{\rm {GeV}}.$
\end{enumerate}
{\small
\begin{table}[htb]
\begin{center}
\begin{tabular}{|c||c|c|c|c|c|c|}
\hline

Process & $\sigma$ (no BRs) & Events after   & Add          & Add      & $\epsilon$ & Events after               \\ 
        &      [pb]         & cuts 1a.--3a.     & $E_T^\ell>100$ GeV 
& $M_T^{1\ell}>200$ GeV &            & ${\cal L}$=100 fb$^{-1}$ \\ 
\hline 
$WH$
%[${\tt IPROC}=3310$]
                     &            1.2 & 116569 & 14101 & 13030 &             0.013 &   1564 \\ \hline
$Z     H$
%[${\tt IPROC}=3360$]
                     &            .69 &   9148 &   794 &   702 &            0.00070 &    48 \\ \hline
$WW$
%[${\tt IPROC}=2800$]
                     &            64. &  38635 &   334 &   235 &            0.00024 &  1504 \\ \hline
$Z     Z$
%[${\tt IPROC}=2810$]
                     &            10. &   4677 &   288 &   253 &            0.00025 &   253 \\ \hline
$W Z$
%[${\tt IPROC}=2820$]
                     &            26. &  32771 &  1180 &  1049 &             0.0010 &  2727 \\ \hline
$W$
%[${\tt IPROC}=1499$]
                     &$1.4\times10^5$ &  81118 &    28 &    24 & $2.4\times10^{-5}$ &336000 \\ \hline
$Z      $
%[${\tt IPROC}=1399$]
                     &$7.5\times10^4$ &    280 &     1 &     0 &                  0 &     0 \\ \hline
$t \bar t$
%[${\tt IPROC}=1706$]
                     &           441. &    661 &    55 &    41 & $4.1\times10^{-5}$ &  1808 \\ \hline
$tq$ + c.c.
%[${\tt IPROC}=2000$]
                     &           146. &   9854 &    10 &     0 &                  0 &     0 \\ \hline\hline
$W+$~jet
%[${\tt IPROC}=2100$]
                     &$6.5\times10^4$ &  70127 &     1 &     0 &                  0 &     0 \\ \hline
$Z    +$~jet
%[${\tt IPROC}=2150$] 
                     &$2.2\times10^4$ &    619 &     0 &     0 &                  0 &     0 \\ \hline
\end{tabular}
\caption{\small Results of the {\tt HERWIG} simulation for the single lepton
channel. The first column gives the normalisation of the hard scattering
processes. The second shows the number of events, out of the $10^6$ generated
in each case, that  survive our preliminary acceptance requirements in 1a.--3a.
The following two columns show the numbers of events surviving
the `sequential' application of the additional cuts on $E_T^\ell$ and
$M_T^{1 \ell}$.  The next column gives  the
overall efficiency of our selection while the last one presents the
final number of events for a luminosity of 
100 fb$^{-1}$.
Note that $Z+$~jet and $W+$~jet are already included and better 
emulated in $Z$ and $W$
production, which in {\tt HERWIG} 
includes the $Z+$~jet 
and $W+$~jet matrix element corrections by default. 
The last two lines are presented for illustrative
purposes only and will not be used in the following for the 
estimation of the signal significance.}
\end{center}
\label{tab1}
\end{table}
}
\normalsize
Tab.~1 summarises our results for the single lepton channel, 
assuming ${\mathrm{{BR}_{~inv}}} = 1$ and $M_H = 120$ GeV, 
coming from both $W H $ and $ZH$ production.  One sees from the table 
that, while all other backgrounds can in the end be reduced to manageable 
level by our sequence of cuts (including those in $E_T^\ell$ and 
$M_T^{1 \ell}$), the background due to off-shell $W^{*}$ production and 
its leptonic decay overwhelms the signal by a factor of more than 200 ! 
Thus the single lepton channel is clearly of little use in the invisible 
Higgs signal extraction. This is  confirmed by the $E_T^\ell$ spectrum 
for the
signal and the leading background  shown in Fig. 1 for the 
luminosity 100~fb$^{-1}$. 
We do not give the 
the $M_T^{1 \ell}$ spectrum as it is strongly correlated to the
one in $E_T^\ell$.
\begin{figure}[htb]
\centerline{
\includegraphics*[scale=0.6]{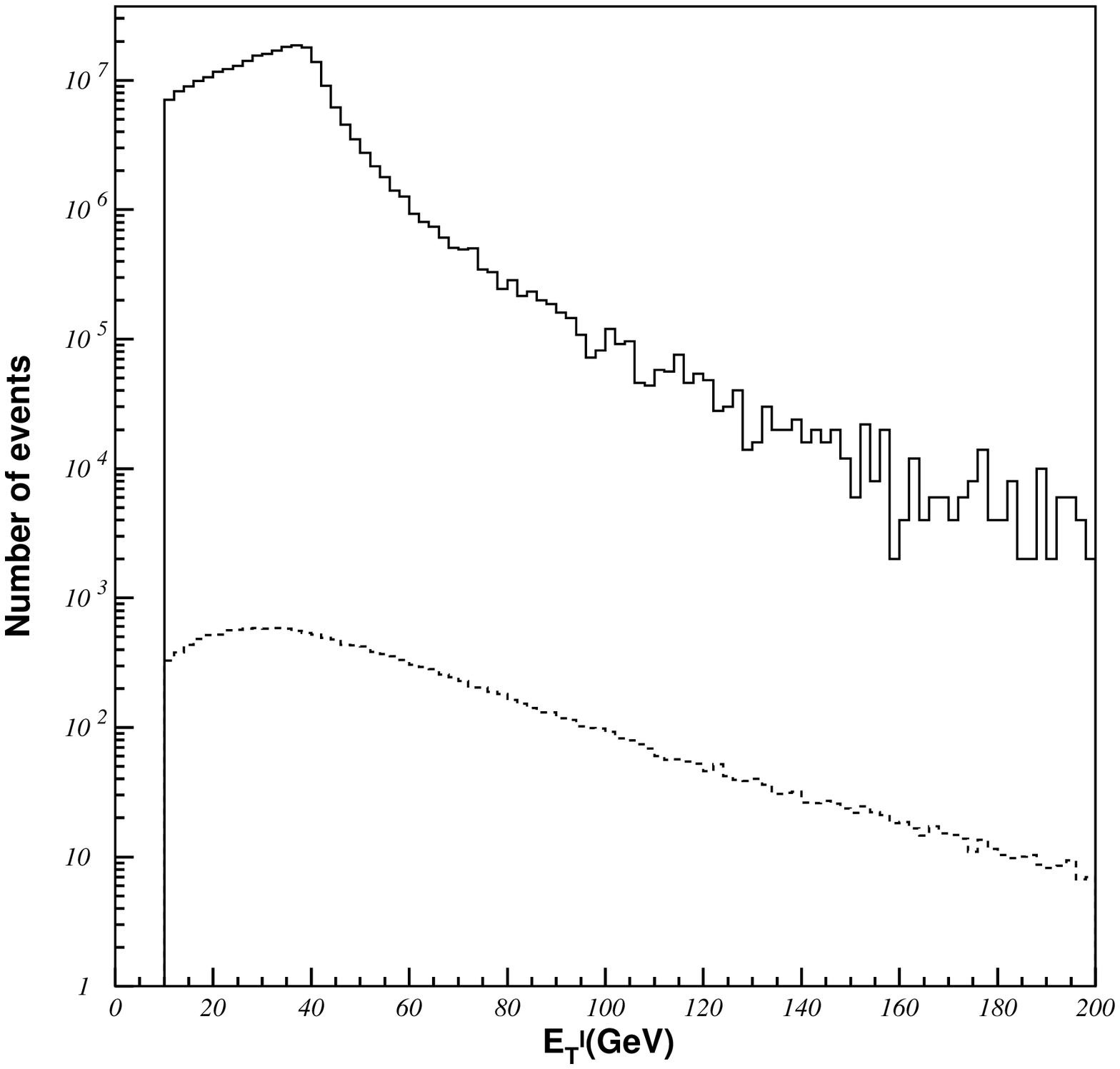}}
\caption{The $E_T^\ell$ distribution for the signal (dashed histogram) and
the dominant charged DY background (solid histogram) in the
case of the one-lepton signature.\label{fig1}}
\end{figure}

For the $\ell \bar \ell+\pt$ channel the situation is much better, 
in spite of the lower signal rates which one starts with. In this case, 
the preliminary acceptance requirements are as follows.
\begin{enumerate}
\item[1b.] Select
 events with  exactly two leptons, same flavour and opposite sign, fulfilling
the kinematic requirements: $|M_{\ell\bar\ell}-M_Z|<10$
GeV, $E_T^\ell > 10~{\rm {GeV}} $ and $ |\eta^\ell| < 3.$

\item[2b.] Impose a hadronic veto, by rejecting events containing jets with
$E_T^j > 30~{\rm {GeV}} $ and $ |\eta^j| < 4.$

\item[3b.]  Enforce a missing transverse momentum threshold:
$\pt >30~{\rm {GeV}}.$
\end{enumerate}
{\small
\begin{table}[htb]
\begin{center}
\begin{tabular}{|c||c|c|c|c|c|c|}
\hline

Process & $\sigma$ (no BRs) & Events after   & Add                & Add           & $\epsilon$ & Events after               \\
        &      [pb]         & cuts 1b.--3b.     & $\pt >100$ GeV & $M_T^{2 \ell}>200$ GeV &            & ${\cal L}$=100 fb$^{-1}$ \\
\hline
$W H$
%[${\tt IPROC}=3310$]
                     &            1.2 & 3 & 0 & 0 & 0 & 0 \\ \hline
$Z     H$
%[${\tt IPROC}=3360$]
                     &            .69 & 28811 & 9593 & 9016 & 0.0090 & 622 \\ \hline
$WW$
%[${\tt IPROC}=2800$]
                     &            64. & 1160 & 16 & 13 & $1.3\times10^{-5}$ & 83 \\ \hline
$Z     Z$
%[${\tt IPROC}=2810$]
                     &            10. & 10618 & 1745 & 1606 & 0.0016 & 1606 \\ \hline
$W Z$
%[${\tt IPROC}=2820$]
                     &            26. & 3374 & 308 & 266 & 0.00026 & 692 \\ \hline
$W$
%[${\tt IPROC}=1499$]
                     &$1.4\times10^5$ & 2 & 0 & 0 & 0 & 0 \\ \hline
$Z      $
%[${\tt IPROC}=1399$]
                     &$7.5\times10^4$ & 6 & 0 & 0 & 0 & 0 \\ \hline
$t \bar t$
%[${\tt IPROC}=1706$]
                     &           441. & 69 & 13 & 9 & $9.0\times10^{-6}$ & 397 \\ \hline
$tq$ + c.c.
%[${\tt IPROC}=2000$]
                     &           146. & 62 & 0 & 0 & 0 & 0 \\ \hline\hline
$W+$~jet
%[${\tt IPROC}=2100$]
                     &$6.5\times10^4$ & 2 & 0 & 0 & 0 & 0 \\ \hline
$Z    +$~jet
%[${\tt IPROC}=2150$]
                     &$2.2\times10^4$ & 16 & 0 & 0 & 0 & 0 \\ \hline
\end{tabular}
\caption{\small Like in Tab. 1 but for the double lepton channel, 
upon replacing the cuts in 1a.--3a. with those in 1b.--3b. and that on 
$E_T^\ell$ with $\pt$.}
\end{center}
\label{tab2}
\end{table}
}
\normalsize

The additional selection cuts here are in  $\pt$ (rather than $E^\ell_T$)
and  $M_T^{2 \ell}$. The former is increased to 100 GeV  while the latter 
is maintained at 200 GeV like  the cut on $M_T^{1\ell}$ for the 
single lepton channel. 
\begin{figure}[htb]
\centerline{
\includegraphics*[scale=0.6]{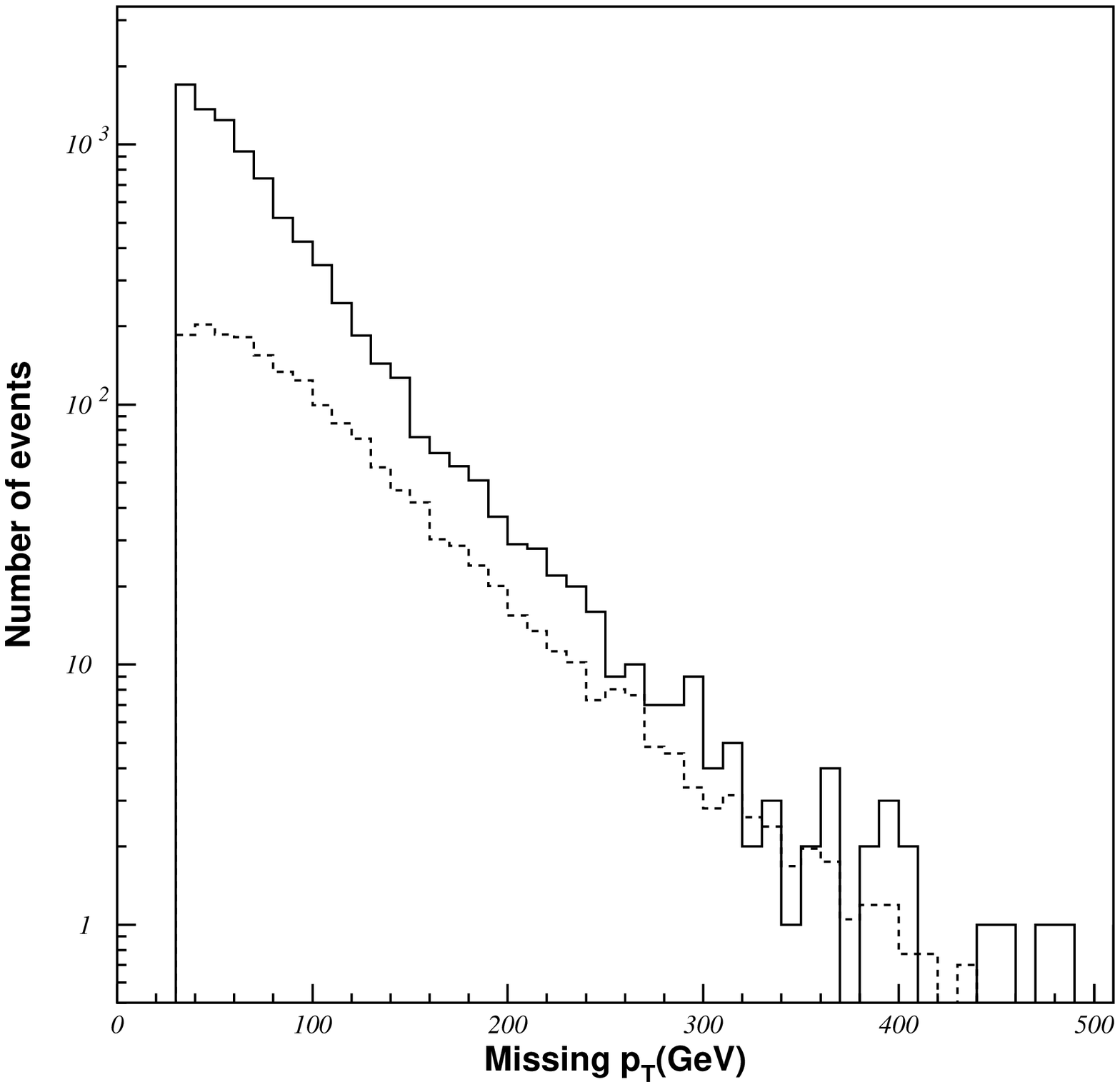}}
\caption{The $\pt$ distribution for the signal (dashed histogram) and
the dominant $ZZ$ background (solid histogram) in the case of the
two-lepton signature.
\label{fig2}}
\end{figure}
Fig. 2 shows the $\pt$ distribution 
for the signal and the leading background for the luminosity 100 fb$^{-1}$. 
Again, we avoid
plotting the $M_T^{2 \ell}$ spectrum as it is very much correlated to the
one in $\pt$ and does not bring any further insights into the kinematics. 
The final results for the dilepton channel, 
for ${\mathrm{{BR}_{~inv}}}=1$ and $M_H = 120$ GeV, 
are summarised in Tab. 2. 
We find that even in this case the signal is surpassed by the background, 
specifically by the $ZZ$ irreducible one and, to a somewhat lesser extent,
by $W Z$ production. 
However both the $ZZ$ and $WZ$ cross--sections are expected to be measured
at LHC to a very good precision via the 4$\ell$ and 3$\ell + \pt$ channels
respectively with $Z$ and $W$ mass reconstructions. 
Since these 
channels should be background free the accuracy of the measured cross
--section will be determined by statistics, i.e about 1\% for the luminosity
of 100~fb$^{-1}$.
Similarly the
$t \bar t$ cross--section is also expected to be measured to a very good
precision. Hence the uncertainty in the number of background events
should be dominated by the statistical fluctuation. 
Moreover, here one can confidently
extract a signal excess, by simply counting the number of dilepton events 
surviving our cuts.  From Tab. 2, the total background cross section 
turns out 
to be 27 fb whereas the signal rate is 6.22 fb for $M_H=120$ GeV, thus 
yielding $S/\sqrt{B} \simeq$7(12) for ${\cal L}=$30(100) fb$^{-1}$  
with ${\mathrm{{BR}_{~inv}}}=$1.

\begin{table}[!t]
\begin{center}
{\small
\begin{tabular}{|c|c|c|}
\hline
$M_H$ & $n_S$ & ${\mathrm{{BR}_{~inv}}}$  \\
(GeV) & ($\#$ events) & $[{\rm {minimum}}$ value] \\
\hline
120 & 187(622) & 0.77(0.42) \\
130 & 158(528)  & 0.88(0.49) \\
140 & 139(462)  & $--$(0.55) \\
150 & 122(407) & $--$(0.64)   \\
160 & 110(366) & $--$(0.70)   \\
\hline
\end{tabular}
}
\caption{\small The 5$\sigma$ discovery limit on the BR of a Higgs boson
decaying invisibly in $ZH$ production, in the dilepton + $\pt$ channel, for 
two LHC luminosities, 30(100) fb$^{-1}$, along with the total number of 
signal events ($n_S$), with the cuts in 1b--3b and the 
additional ones mentioned 
in the caption of Tab. 2.} 
\end{center}
\end{table}

These numbers are promising enough to further investigate the chances 
of extracting a signal for other combinations of ${\mathrm{{BR}_{~inv}}}$
and $M_H$. In Tab. 3, we present the lower limits on 
${\mathrm{{BR}_{~inv}}}$ for which a $\gsim5\sigma$ excess is possible in 
the dilepton channel for an IMH boson, for two different values of 
integrated luminosities, 30 and 100 fb$^{-1}$.  
We see that with a luminosity of 100 fb$^{-1}$ a discovery is possible down to
${\mathrm{{BR}_{~inv}}} =0.42$ for $M_H$ = 120 GeV and to 
${\mathrm{{BR}_{~inv}}} = 0.7$ for $M_H$ = 160 GeV.
It should be mentioned here that, even for ${\cal L} = 30$
fb$^{-1}$, a $\sim4\sigma$ level signal is possible up to $M_H
=160$ GeV for ${\mathrm{{BR}_{~inv}}} = 1$.
Notice all the ${\mathrm{{BR}_{~inv}}}$ values 
discussed here are consistent with current LEP limits on 
`$H\to$~invisible' processes~\cite{LHWG}.

Finally note that the differential shape of the signals and backgrounds 
have become very similar after our selection cuts and hence one
should expect only a limited margin of improvement on the rates presented
here, from the application of any further kinematic constraints\footnote{Our 
results for both the channels are consistent with those obtained by the  
ATLAS study in \cite{vhsim}.}.

\section*{Conclusions}

In summary, we have studied possibility of Higgs detection in
invisible channels at LHC, via the production mode $q\bar q\to ZH$,
followed by leptonic decays of the gauge boson using electrons and/or
muons in the final state. The signature arises in the form of an excess 
in the total number of `dilepton plus missing transverse energy' events. 
The channel is viable over the entire intermediate mass interval
114 GeV $\lsim M_H\lsim$ 160 GeV for an accumulated luminosity of 100 fb$^{-1}$
and the viability is limited to  $M_H\lsim$ 130 GeV if the available 
luminosity is also limited to 30 fb$^{-1}$.

Should the reach in the traditional $b\bar b$, $\gamma\gamma$ and $VV^*$
detection modes be diminished for a relatively light Higgs boson
due to  its novel invisible decays, one sees a reasonable chance to 
detect the latter, which may thus help to compensate for the suppression 
of the former. We have demonstrated  this by performing a rather 
detailed and largely model independent MC study at hadron level and 
in presence of detector effects. Our results call
for a combined analysis of the $H\to b\bar b, \gamma\gamma, VV^*$ and `invisible'
modes to establish LHC potential to discover an intermediate mass 
Higgs boson even in the presence of substantial partial decay width into 
the invisible mode.

\section*{Acknowledgements}
SM is grateful to CERN Geneva and IPPP Durham for the use
of their computing resources in finalising this work.
RG and MG acknowledge the hospitality of the Theory Division at CERN, where
a major part of this work was completed.We acknowledge useful discussions 
with Sunanda Banerjee, M. Dittmar, B. A. Kniehl and A. Nikitenko.
KM acknowledges CMS computing facilities
where part of the simulations were made.
We wish to thank  
the organisers of the Workshop on High Energy Physics Phenomenology (WHEPP-7),
held at Harish Chandra Research Institute, Allahabad, India, in January 2002,
where this project was initiated.

\end{document}